\documentstyle[epsf,eqsecnum,floats,preprint,aps]{revtex}

\input epsf
\newcommand{\PSbox}[3]{\mbox{\rule{0in}{#3}\includegraphics{#1}\hspace{#2}}}

\def\balpha{\mbox{\boldmath $\alpha$}}
\def\bbeta{\mbox{\boldmath $\beta$}}

\def\bgamma{\mbox{\boldmath $\gamma$}}
\def\bmu{\mbox{\boldmath $\mu$}}
\def\bnu{\mbox{\boldmath $\nu$}}

\begin{document}

\rightline{CU-TP-1024}
\rightline{hep-th/0108029}
\vskip 1cm
\baselineskip 18pt

\begin{center}
\ \\
\large{{\bf Massless monopoles and the moduli space approximation }} 
\ \\
\ \\
\normalsize{Xingang Chen,\footnote{Email address: \tt xgchen@phys.columbia.edu}
Huidong Guo,\footnote{Email address: \tt guohd@phys.columbia.edu} and  Erick
J. Weinberg\footnote{Email address: \tt ejw@phys.columbia.edu}}   
\ \\
\small{\em Department of Physics, 
Columbia University, 
New York, NY 10027}

\end{center}
\begin{abstract}
\baselineskip=18pt

  We investigate the applicability of the moduli space approximation in
theories with unbroken non-Abelian gauge symmetries.  Such theories
have massless magnetic monopoles that are manifested at the classical
level as clouds of non-Abelian field surrounding one or more massive
monopoles.  Using an SO(5) example with one massive and one massless
monopole, we compare the predictions of the moduli space
approximation with the results of a numerical solution of the full
field equations.  We find that the two diverge when the cloud velocity
becomes of order unity.  After this time the cloud profile
approximates a spherical wavefront moving at the speed of light.  In
the region well behind this wavefront the moduli space approximation
continues to give a good approximation to the fields.  We therefore 
expect it to provide a good description of the motion of the
massive monopoles and of the transfer of energy between the massive
and massless monopoles.

\end{abstract}

\setcounter{page}{0}
\thispagestyle{empty}
\maketitle
\baselineskip=18pt

\section{introduction}

The moduli space approximation (MSA) \cite{Manton:1982mp} is a useful
tool for studying the low-energy dynamics of
Bogomolny-Prasad-Sommerfield (BPS) magnetic monopoles
\cite{Bogomolny:1976de,Prasad:1975kr} and other systems that allow
families of static multisoliton solutions.  In this approximation the
full field dynamics is replaced by that of the small number of
collective coordinates that span the moduli space of static solutions.
This neglects possible distortions of the inner structure of the
solitons, and thus effectively treats them as point particles.  The
approximation is expected to be valid when the soliton velocity $v\ll
1$, with corrections suppressed by powers of $v$.  This suggests that
it might break down in certain gauge theories where, as we describe
below, some BPS monopoles are effectively transformed into massless
particles \cite{Lee:1996vz}.  Indeed, a study by two of us
\cite{Chen:2001ge} of monopole scattering in one such theory
encountered anomalies that suggest a breakdown of the MSA.  In this
paper we describe some analytic and numerical calculations that shed
light on this issue.

The MSA can be motivated by recalling some features of the spectrum of
small oscillations about a multisoliton solution.  First, there are a
number of zero-frequency modes that are handled by the introduction of
the collective coordinates.
Next, for each elementary field of mass $m_i$ in the theory there is
a continuum spectrum beginning with minimum energy $E=m_i$.  Finally,
there may be some discrete nonzero eigenvalues.  The last two
components of the spectrum correspond to scattering states and bound
states, respectively, of the massive quanta in the presence of the
solitons.  If the soliton kinetic energies are much smaller than the
$m_i$, excitation of these modes should be suppressed.  In the MSA one
assumes that this suppression is complete and takes the collective
coordinates to be the only dynamical degrees of freedom.  If, as is
the case for BPS solutions, the static energy is
independent of the collective coordinates, the effective Lagrangian
for the collective coordinates has only kinetic energy terms.  These
define a metric on the moduli space, and the classical evolution of
the system is equivalent to geodesic motion with respect to this
metric.

The energetic arguments underlying the MSA apply fairly
straightforwardly to theories with only massive fields.  However, when
massless fields are present, as is the case in theories with magnetic
monopoles, radiation of low-energy quanta is always energetically
possible.  To justify the MSA, one must show that this radiation is
suppressed at low monopole velocities.  The simplest case to consider
is that of an SU(2) theory broken to U(1), with gauge coupling $e$, a
massive vector boson of mass $m$, and a monopole mass $M \sim 4\pi
m/e^2$.  The problem of radiation in this theory was first addressed
by Manton and Samols \cite{Manton:1988bn}.  For well-separated
monopoles the amount of U(1) ``electromagnetic''
radiation\footnote{There is also radiation of the massless scalar
field, which can be handled by similar methods.} can be estimated by
treating the monopoles as point magnetic charges moving along the
trajectories specified by the MSA.  Standard electromagnetic
techniques show that the total dipole radiation is of order $Mv^3$,
while higher multipoles are suppressed by additional powers of $v$.
(For the case of two monopoles, the dipole moment vanishes
identically, so the quadrupole term dominates and the total radiation
is of order $M v^5$.)  These arguments break down if the monopoles
approach closely enough that their cores actually collide and overlap.
However, since the core radius is of order $m^{-1}$, the dominant
effect should only be on modes of wavelength $\lesssim m^{-1}$, with
quanta of energy $\gtrsim m$, and these can neglected because of
energetic considerations.  These somewhat heuristic arguments have
been placed on a more rigorous footing by Stuart \cite{Stuart:1994tc}.

Now consider BPS monopoles in a theory with a gauge group $G$
of rank $k>1$.  For generic vacuum expectation values of the adjoint
Higgs field the symmetry is broken to the maximum Abelian subgroup
U(1)$^k$.  There are then $k$ distinct topological charges.   Associated with
each is a fundamental monopole \cite{Weinberg:1980zt}, carrying one
unit of that charge, that 
can be realized as an embedding of the SU(2) unit monopole in the SU(2)
subgroup defined by one of the simple roots of $G$.  All higher-charged
BPS solutions can be understood as multimonopole solutions containing
appropriate numbers of the various species of fundamental monopoles.  As
long as the fundamental monopoles all remain massive, the arguments used
to justify the MSA in the SU(2) theory can be readily extended to this more
general case.

The situation is more complex if the Higgs field is such that the
unbroken gauge group has a non-Abelian component.  When this happens,
the masses of some of the fundamental monopoles vanish.
These massless monopoles, which can be viewed as the counterparts of the
massless non-Abelian gauge bosons, cannot be realized as isolated
classical solutions.  Instead, they are manifested in classical
multimonopole solutions as ``clouds'' surrounding one or more massive
monopoles \cite{Lee:1996vz}.  Both Abelian and non-Abelian magnetic
fields are present 
within the cloud region, but outside the cloud the non-Abelian
components are suppressed by powers of $r$ relative to the Abelian
Coulomb magnetic fields.  For static solutions the energy is
independent of the size of the cloud.

One might well expect to encounter some difficulties with the MSA in
the case.  First, the presence of massless non-Abelian gauge fields gives
more possibilities for low-energy radiation.  Second, the charged fields
that are the source for this radiation are not confined to a core
region of size $\sim m^{-1}$, as in the case of Abelian breaking, but
are instead present throughout the cloud region, which can be
arbitrarily large.  Finally, the vanishing mass of some of the
monopoles seems inconsistent with the low-velocity regime that is
required for the MSA.

This last point needs some elaboration.  Because the massless
monopoles do not have well-defined positions, their velocities are
also not well-defined.  However, there is an associated quantity with
dimensions of velocity, namely the rate of expansion of the cloud
region.  This need not be unity, as an ordinary massless-particle
velocity would be.  On the other hand, even if this ``cloud velocity''
is initially small, there is no guarantee that it will remain small,
or even bounded.  For example, in Ref.~\cite{Chen:2001ge}, the MSA was
used to study solutions containing two massive monopoles and a
massless cloud in an SU($N$) theory.  It was found that the massive
monopoles tend toward a constant velocity at large times, but that the
cloud size varies roughly as $kt^2$, where $k$ is proportional to the
kinetic energy associated with the cloud.  Thus, no matter how small
this energy, the MSA predicts that the cloud eventually expands
faster than the speed of light.

The prediction of such superluminal expansion in a relativistic field
theory suggests a breakdown of the MSA.  In principle, one could test
this by solving the full time-dependent field equations and comparing
with the MSA predictions.  For arbitrary configurations of the SU($N$)
monopoles this would be a fairly difficult undertaking.  We therefore
focus instead on another example that shares many of the essential
features of the SU($N$) solutions but that is analytically more
tractable.

The example \cite{Weinberg:1982jh} we consider occurs in a theory with
an SO(5) gauge symmetry broken to SU(2)$\times$U(1) by an adjoint
Higgs field.  There are two species of fundamental monopoles, one
massive and one massless.  The static BPS solutions containing one of
each are all spherically symmetric and contain a massive core of fixed
radius surrounded by a cloud of arbitrary radius $a$.  Suppressing the
dependence on the other collective coordinates, we denote the
corresponding fields by $A_\mu^{\rm BPS}({\bf r}, a)$ and $\Phi^{\rm
BPS}({\bf r}, a)$.

If we work in the center-of-mass frame and restrict ourselves to
solutions with vanishing SU(2) and U(1) electric charges, the cloud
radius is the only time-dependent collective coordinate.  Solving the
Euler-Lagrange equation following from the moduli space Lagrangian
gives a solution $a_{\rm MSA}(t)$. The MSA consists of assuming that
the time-dependence of the fields arises solely from that of $a$, and
that the latter is given by $a_{\rm MSA}(t)$.  Explicitly,
\begin{eqnarray}
   A_\mu^{\rm MSA}({\bf r}, t) &=& A_\mu^{\rm BPS}({\bf r}, a_{\rm
   MSA}(t)) \, ,  \cr 
   \Phi^{\rm MSA}({\bf r}, t) &=& \Phi^{\rm BPS}({\bf r}, a_{\rm MSA}(t))
     \, . 
\label{MSAdef}
\end{eqnarray}

To test this approximation, we start with a configuration that is of
this form at $t=0$; i.e., that satisfies 
\begin{eqnarray}
   A_\mu({\bf r}, 0) &=& A_\mu^{\rm BPS}({\bf r}, a_0)  \, , \cr\cr
   \Phi({\bf r}, 0) &=& \Phi^{\rm BPS}({\bf r}, a_0) \, , \cr \cr
  \dot A_\mu({\bf r}, 0) &=& \dot a_0 
    {\partial A_\mu^{\rm BPS}({\bf r}, a_0)  \over\partial a} \, , \cr\cr
  \dot \Phi({\bf r}, 0) &=& \dot a_0 
    {\partial \Phi^{\rm BPS}({\bf r}, a_0)  \over\partial a} \, ,
\label{MSAinitcond}
\end{eqnarray}
with dots denoting time derivatives.
Since we expect the MSA to be most reliable when velocities are small,
we take the initial cloud velocity $\dot a_0 \ll 1$.  We also
assume that the initial cloud size $a_0$
is much greater than the radius of the massive core.  

These initial conditions determine a solution of the full field
equations that can be compared with the MSA solution of
Eq.~(\ref{MSAdef}).  Although the spherical symmetry implied by the
initial conditions simplifies the field equations somewhat, one is
still left with a large number of coupled differential equations.
However, the fact that the departures from the MSA are expected to
occur primarily outside the massive core leads to a further simplification
that reduces the problem to a single second-order equation.
Analysis of this equation leads to a semi-quantitative understanding
of the breakdown of the MSA.   To test these ideas in detail, however, 
numerical methods are required.

In Sec.~II we review the SO(5) theory in detail and describe the
moduli space of BPS solutions.  In
Sec.~III we set up the initial value problem that we will be
considering.  Here we analyze the field equations and obtain
predictions for the expected departures from the MSA solution.  In
Sec.~IV we describe the numerical solution of the field
equations, and compare it with our expectations.  In Sec.~V we
review our results and discuss their implications for the use of the
MSA.

\section{SO(5) monopoles}

\begin{figure}[hbtp]\begin{center}
\PSbox{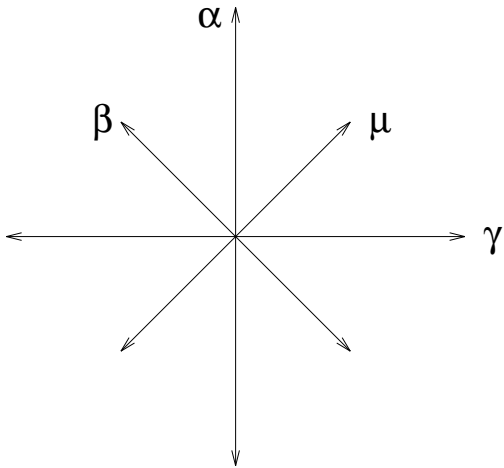
hscale=80 vscale=80  hoffset=20    
voffset=-65}{3.0truein}{2.0truein}
\end{center} 
\medskip
\caption{The root diagram of SO(5)}
\label{fig:so5roots}
\end{figure}

The generators of SO(5) may be chosen to be two commuting operators
$H_1$ and $H_2$ that generate the Cartan subalgebra, together with
eight raising and lowering operators $E_{\bnu_j}$, where the $\bnu_j$
are the roots shown in Fig.~\ref{fig:so5roots}.  (We fix the
normalization of the gauge coupling by taking the long roots to be of
unit length.)  The vacuum expectation value of the adjoint Higgs field
can always be chosen to lie in the Cartan subalgebra, and hence to be
of the form $\Phi_\infty = h_1 H_1 + h_2 H_2 \equiv {\bf h}\cdot {\bf
H}$.  Furthermore, $\bf h$ can always be chosen so that ${\bf h}\cdot
\bbeta$ and ${\bf h}\cdot \bgamma$ are both positive.  The two species
of fundamental monopoles are then associated with the roots $\bbeta$
and $\bgamma$ and have masses
\begin{equation} 
    M_{\bbeta} = {4\pi \over e} {{\bf h}\cdot \bbeta\over \bbeta^2}
     = {8\pi \over e}{\bf h}\cdot \bbeta \, ,
\qquad\qquad   M_{\bgamma} = {4\pi \over e} {{\bf h}\cdot  \bgamma\over
     \bgamma^2}  = {4\pi \over e}{\bf h}\cdot  \bgamma \, .
\end{equation}
The corresponding electrically-charged elementary gauge bosons have
masses 
\begin{equation}
    m_{\bbeta} = e {{\bf h}\cdot \bbeta} \, ,
\qquad\qquad   m_{\bgamma} = e {{\bf h}\cdot  \bgamma} \, .
\end{equation}
If ${\bf h}\cdot \bbeta$ and ${\bf h}\cdot \bgamma$ are both nonzero,
the unbroken gauge group is U(1)$\times$U(1).  If instead $\bf h$ is
orthogonal to $\bgamma$, the symmetry is only broken to
SU(2)$\times$U(1), and the $\bgamma$-monopole mass, like $m_{\bgamma}$,
vanishes; this is the case we wish to consider.  For convenience, we
will define $v = {\bf h}\cdot \balpha = 2 {\bf h}\cdot \bbeta$. 

Index theorems \cite{Weinberg:1982ev} show that there is an
eight-dimensional moduli space of BPS solutions containing one massive
$\bbeta$- and one massless $\bgamma$-monopole.  Three of the
collective coordinates can be chosen to be the center-of-mass
coordinates $\bf X$, while four others can be taken to be the angles
$\alpha$, $\beta$, $\gamma$, and $\chi$ that specify the global
SU(2)$\times$U(1) orientation of the solution.  Variations of these
seven variables are equivalent to the actions of symmetry
transformations.  The dependence of the solutions on the last
collective coordinate, which we take to be the cloud parameter $a$, is
less trivial.

To display this dependence, we need some notation.  We assemble the
generators into a triplet
\begin{eqnarray}
   t_1(\balpha) &=& {1 \over \sqrt{2}}(E_{\balpha} + E_{-\balpha} ) 
      \, ,\cr\cr
   t_2(\balpha) &=& - {i \over \sqrt{2}}(E_{\balpha} - E_{-\balpha} )
   \, , \cr\cr 
   t_3(\balpha) &=&  \balpha \cdot {\bf H} \, ;
\end{eqnarray}
a second triplet $t_a(\bgamma)$, defined by analogous expressions
involving $\bgamma$; and a quartet assembled
into a $2\times 2$ matrix
\begin{equation}
    M = \sqrt{2} \, i\left(\matrix{E_{\bbeta}  &-E_{-\bmu} \cr 
       E_{\bmu}   &E_{-\bbeta}  }\right) \, .
\end{equation}
The spacetime fields can then be written as
\begin{eqnarray}
A_{i}&=&{\bf A}_i^{(1)} \cdot {\bf t}({\bf \alpha}) + {\bf A}_i^{(2)}
\cdot {\bf t}({\bf \gamma}) + {\rm tr}\,{\bf A}_i^{(3)} M \, ,
\nonumber\\
\Phi&=&{\bf \Phi}^{(1)} \cdot {\bf t}({\bf \alpha}) + {\bf \Phi}^{(2)}
\cdot {\bf t}({\bf \gamma}) + {\rm tr}\,{\bf \Phi}^{(3)} M  \, .
\label{so5ansatz}
\end{eqnarray}
With the ${\rm SU}(2) \times {\rm U}(1)$ symmetry breaking that we are
assuming, the asymptotic Higgs field lies in the subgroup generated by
the ${\bf t}(\balpha)$.  The field components labeled by superscripts
1, 2, and 3 transform as three singlets, a triplet, and a complex
doublet, respectively, under the unbroken SU(2).

In Ref.~\cite{Weinberg:1982jh} it was shown that there is a static
solution of the form 
\begin{eqnarray}
A^a_{i(1)} &=& \epsilon_{aim} {\hat r}_m A(r)\, , \qquad
\Phi^a_{(1)}={\hat r}_a H(r)\, ,\nonumber\\
A^a_{i(2)} &=& \epsilon_{aim} {\hat r}_m G(r)\, , \qquad
\Phi^a_{(2)}={\hat r}_a G(r)\, ,\nonumber\\
A^a_{i(3)} &=& \tau_i F(r)  \, , \qquad \qquad \,    \Phi_{(3)}=-iIF(r) \, .
\label{solutionbegin}
\end{eqnarray} 
Here 
\begin{equation}
A(r) = \frac{v}{\sinh evr} - \frac{1}{er}
\end{equation}
and
\begin{equation}
H(r) = v\coth evr - \frac{1}{er}
\end{equation}
are the functions appearing in the SU(2) unit BPS monopole, while 
\begin{equation}
  F(r)=\frac{v}{\sqrt 8 \cosh (evr/2)} L(r,a)^{1/2}
\end{equation}
and
\begin{equation}
  G(r) =A(r)L(r,a) \, ,
\label{staticG}
\end{equation}
with
\begin{equation}
L(r,a)=[1+(r/a)\coth (evr/2)]^{-1} \, .
    \label{solutionend}
\end{equation}

At the center of this solution is a massive core of radius $\sim
1/ev$.  Surrounding this is a non-Abelian cloud region
whose size is set by the parameter $a$, which can take on any positive
real value.  Inside the cloud, in the region $1/ev \ll r \ll a$, the
non-Abelian vector potential ${\bf A}_i^{(2)}$ falls as $1/r$ and
yields a Coulomb non-Abelian magnetic field.  In the
region $r \gg a$, the non-Abelian magnetic charge is canceled by the
cloud, and ${\bf A}_i^{(2)}$ falls more rapidly, as $1/r^2$.  

Given the explicit form of the solution, it is a straightforward
matter to obtain the metric on the moduli space, and thus the moduli
space Lagrangian \cite{Lee:1996vz}
\begin{equation}
    L_{\rm MS} = {1\over 2} M_{\balpha} \dot {\bf X}^2 
       + {8\pi^2 \over M_{\balpha} e^4} \dot\chi^2  
      + {2 \pi \over e^2 } \left\{ {\dot a^2 \over a} + a\left[
      \dot \alpha^2 + \sin^2\alpha \,\dot\beta^2 +
      (\dot\gamma + \cos\alpha \,\dot\beta)^2 \right] \right\}  \, .
\label{SO5msaLag}
\end{equation}

\section{Field equations and analytical predictions}

Our goal in this paper is to understand how well the solutions given
by the MSA approximate the solutions of the full field equations.  To
simplify matters, we will restrict ourselves to solutions with
vanishing linear momentum and SU(2)$\times$U(1) charges.  The
Euler-Lagrange equations that follow from Eq.~(\ref{SO5msaLag}) then
imply that the only time-dependent collective coordinate is $a$, given by
\begin{equation}
     a_{\rm MSA}(t) \equiv k(t+\Delta)^2 
\label{MSAcloudsize}
\end{equation}
where $k$ and $\Delta$ are constants.  The associated kinetic energy
is
\begin{equation}
   E =  {2 \pi \over e^2 }{\dot a^2 \over a} 
     =  {8 \pi k\over e^2 }\, .
\label{MSAenergy}
\end{equation}

The MSA approximation to the fields is obtained by substituting
Eq.~(\ref{MSAcloudsize}) into Eq.~(\ref{MSAdef}) We want to compare
this with the solution of the full field equations.  Because of the
spherical symmetry of the static BPS solutions, our initial conditions
imply that the solutions will be spherically symmetric for all $t$.
Even with this taken into account, the most general Ansatz for 
the fields would involve a large number of functions of $r$ and $t$.
However, we expect that the deviations from the MSA arise primarily
from the dynamics of the massless fields, rather than from the massive
fields that are confined to the monopole core.  Furthermore, we expect
these deviations to manifest themselves primarily at large distances.
Let us therefore choose some radius $\bar r$ that is large compared to
the monopole core radius, but much smaller than the initial cloud
size; i.e., $1/ev \ll \bar r \ll a_0$.  We will assume that the MSA
gives a good approximation to the fields at $r \ll \bar r$ for all
$t$.  (We will see that our numerical results are consistent with this
assumption.)  We then only need to solve the field equations for $r >
\bar r$, subject to the boundary condition that the solution match
onto the MSA solution at $r=\bar r$.

This leads to considerable simplification.  The massive SU(2) doublet
fields $A_\mu^{(3)}$ and $\Phi^{(3)}$ fall exponentially outside
the monopole core and can therefore, to a good approximation, be set
equal to zero in the region $r > \bar r$.  Once this is done, the
singlet and triplet fields are decoupled from each other.  Since the
former are independent of $a$, we can concentrate on the latter.

Requiring spherical symmetry and positive parity implies that the
triplet fields 
must be of the form
\begin{eqnarray}
    A_i^{a(2)} & =&   \epsilon_{aim} {\hat r}_m [G(r,t)+K(r,t)] \, ,\cr
    A_0^{a(2)} & =&  {\hat r}_a J(r,t)  \, ,\cr
    \Phi^{a(2)} & =& {\hat r}_a [G(r,t)-K(r,t)]\, .
\end{eqnarray}
$J$ can be set equal to zero by
a time-dependent gauge-transformation.  The 
Euler-Lagrange equations of the theory then reduce to 
\begin{equation}
   - \ddot G + G'' + {2 \over r}G' - {2 \over r^2}G 
    - {2e\over r}G(2G+K) - 2e^2(G+K)G^2  = 0
\end{equation}
\begin{equation}
   - \ddot K + K'' + {2 \over r}K' - {2 \over r^2}K 
    - {2 e\over r}K(2K+G) -2e^2(G+K)K^2 = 0
\label{Keq}
\end{equation}
where dots and primes denote differentiation with respect to $t$ and
$r$, respectively, and exponentially small quantities have been ignored. 

The initial conditions of Eq.~(\ref{MSAinitcond}) imply that both $K$
and $\dot K$ vanish for all $r$ at $t=0$.  It then follows from
Eq.~(\ref{Keq}) that $K$ will continue to vanish for all time.  (One
can also verify that small deviations from $K=0$ do not produce
growing perturbations.)
Defining $g(r,t) =- e r G(r,t)$, we can then write
\begin{equation}
 - \ddot g + g''  - {2 \over r^2} g (1- g)^2 =0  \, .
\label{geq}
\end{equation}
This has the static solution
\begin{equation}
     g_{\rm MS}(r,a) = \frac{a}{r+a}
\label{gm}
\end{equation}
which corresponds to the $r\gg 1/ev$ limit of the BPS solution
Eq.~(\ref{staticG}).  Note that $g \approx 1$ far inside the cloud,
while far from the cloud $g$ tends to zero.

The initial conditions of Eq.~(\ref{MSAinitcond}) reduce to 
\begin{eqnarray}
   g(r,t=0) &=& g_{\rm MS}(r,a_0)  \, ,\cr\cr
   \dot g(r,t=0) &=& \dot a_0 \left.{\partial g_{\rm MS}(r,a) \over
   \partial a}\right|_{a=a_0}  =  { r \dot a_0 \over (r + a_0)^2} \, .
\label{ginitcond}
\end{eqnarray}
In addition, we have the boundary condition 
\begin{eqnarray}
    g(\bar r, t) = g_{\rm MS}(\bar r,a_{\rm MSA}(t)) =
    \frac{a_{\rm MSA}(t)}{\bar r+a_{\rm MSA}(t)} \, .
\label{gboundary}
\end{eqnarray}
The time dependence of 
this boundary condition implies that the total energy in the region $r
> \bar r$ will not be conserved.  Physically, this is to be expected.
Although the total energy of the static solutions is independent of
$a$, the division of this energy between the core and the cloud 
varies with cloud size.  Hence, even within the MSA there will be energy
flow across the surface
$r = \bar r$.

We will find it necessary to use numerical methods to solve 
Eq.~(\ref{geq}).  We will describe these in the next section.  Before
doing so, let us consider what type of departures from the MSA we
might expect.
We can distinguish two different ways in which 
the MSA might be expected to fail.  First, the solution may depart from the
moduli space by the excitation of low-frequency modes; i.e., by
radiation.  In the present context, this can occur because the
time-dependent fields implied by the MSA act as a source for
non-Abelian radiation.  To make this more precise, let us write
$g(r,t)=g_{\rm MS}(r,a_{\rm MSA}(t))+\delta g(r,t)$.  If $\delta g \ll g_{\rm
MS}$, we can linearize Eq.~(\ref{geq}) to obtain
\begin{eqnarray}
-\ddot \delta g + \delta g''-\frac{2}{r^2} (1-g_{\rm MS})(1-3 g_{\rm
       MS})\delta g 
       &=&\ddot g_{\rm MS}   \cr\cr
     &=& -{2 \dot a_{\rm MSA}^2 r \over (r+a_{\rm MSA})^3} 
  + {\ddot a_{\rm MSA} r \over (r+a_{\rm MSA})^2} \cr\cr
   &=& {e^2 E \over 4\pi} {r (r-3a_{\rm MSA}) \over (r+a_{\rm MSA})^3} \, .
\label{deltag}
\end{eqnarray}
where $E$ is the cloud kinetic energy given by Eq.~(\ref{MSAenergy})

The right hand side of this equation is concentrated in the cloud
region, $r \lesssim a$.  (For this initial analysis of the problem we
ignore the $1/r$ tail.)  Hence, we can view $\ddot g_{\rm MS}$ as
the source for a radiation field $\delta g$ that propagates outward
from the cloud at the speed of light.  In principle, we could solve
for $\delta g$ in terms of this source by using Green's function
methods.  However, because of the presence of the terms involving
$g_{\rm MS}$, we have not been able to obtain a closed form expression
for the Green's function.

Nevertheless, we can still estimate the rate of radiation; i.e., the
power $P$ that passes through a sphere of some fixed radius $r \gg
a_0$.  From Eq.~(\ref{deltag}) we see that $\delta G = \delta g/(er)$
is proportional to $eE$.  Since the radiation energy density
is quadratic in derivatives of $\delta G$, the power must be 
of the form
\begin{equation}
    P \sim  e^2 E^2 f(\dot a) \sim E {\dot a^2 \over a} f(\dot a)  \, ,
\label{power}
\end{equation}
with dimensional arguments showing that $f$ cannot depend on $a$. It is 
also easy to see that $f$ cannot depend on $e$.  Although further analysis 
would be required to determine the form of $f(\dot a)$ at small $\dot a$,
it is presumably of 
order unity when $\dot a$ is of order unity.

The MSA must certainly break down by the time that the integrated
power becomes comparable to the total
energy of the cloud.  Comparing Eqs.~(\ref{MSAenergy}) and (\ref{power})
and noting that the bulk of the energy loss occurs at large $\dot a$, where 
$f$ is of order unity, we
see that this happens when $t \sim a /\dot a^2 \sim  k^{-1}$.  This
is also the time that Eq.~(\ref{MSAcloudsize}) predicts a cloud velocity
comparable to the speed of light, a second indication of the breakdown
of the MSA.

Because of its finite speed of propagation, the radiation described
above can have no effect on the behavior of the fields at large
distances; i.e., at $r > t$.  However, there is a second type of
breakdown of the MSA that occurs in this region.  The MSA assumes that
the system can be well described by a small number of time-dependent
collective coordinates.  This assumption can only be valid if
retardation effects can be ignored, so that the system effectively
reacts as a single unit to changes in the collective coordinates.
This should be true within the cloud region as long as the
characteristic time $a/\dot a$ is much greater than $a$; i.e., as long
as the cloud velocity is much less than the speed of light.  However,
no matter how small $\dot a$ may be, we should expect departures from
the MSA at distances $r \gg a/\dot a$.

We can examine this in more detail by using a large-distance expansion
to solve Eq.~(\ref{geq}).  Let 
\begin{equation}
   g(r,t) = \sum_{n=1}^\infty {C_n(t) \over r^n} \, .
\label{largeRexp}
\end{equation}
Substituting this expansion into Eq.~(\ref{geq}) gives a series of
equations for the $C_n(t)$.  The first two of these require that
$\ddot C_1 = \ddot C_2 =0$.  Integrating these with the initial
conditions implied by Eq.~(\ref{ginitcond}) gives
\begin{eqnarray}
    C_1 &=& a_0 + \dot a_0 t    \, ,\cr
    C_2 &=& -a^2_0 - 2 a_0\dot a_0 t  \, .
\end{eqnarray}
Hence, 
\begin{equation}
   g(r,t) \approx g_{\rm MS}(r, a_0+\dot a_0 t) \, .
\label{largeRg}
\end{equation}
In other words, the large-distance fields are close to the moduli
space of BPS solutions, but with a cloud size that grows at the
initial velocity $\dot a_0$ rather than at the accelerating velocity
predicted by the MSA.

\section{Numerical results}

We now turn to our numerical results.  As explained in the previous
section, we assume that the MSA is valid within a region of radius
$\bar r \gg 1/ev$, and so only need to obtain a numerical solution for
the region $r > \bar r$.  The field equations then reduce to the
single equation Eq.~(\ref{geq}), which is to be solved subject to the
initial conditions of Eq.~(\ref{ginitcond}) and the boundary
condition of Eq.~(\ref{gboundary}).  To fix the initial conditions we
must specify the initial cloud size, $a_0$, and the initial cloud
velocity, $\dot a_0$.  Because the cloud size sets the characteristic
length scale for the spatial variations of the field, discretization
errors will be reduced if $a_0 \gg \bar r$; for our solutions we take
$a_0= 50 \bar r$.  (The actual numerical value has no direct
physical meaning, since our approximations effectively set $1/ev$, the
only other physical length scale, equal to zero.)  In order that we
initially be in a regime where the MSA can be expected to be valid,
$\dot a_0$ should be nonrelativistic; we choose $\dot a_0 =0.01$.  The
parameters in Eq.~(\ref{MSAcloudsize}) are then $k= 2.5 \times 10^{-5}
a_0^{-1}$ and $\Delta= 200 a_0$.

We expect the MSA to break down at roughly the time that $\dot a_{\rm
MSA}$ becomes of order unity; i.e., at $t \sim t_{\rm crit}$, where
\begin{equation}
       t_{\rm crit} = {1 \over 2k} = 2 \times 10^{4} a_0 \, .
\end{equation}
According to Eq.~(\ref{MSAcloudsize}), $a_{\rm MSA}$ would then be
$\sim 1/(4k) = 10^4 a_0$.  Hence, to see the breakdown of the MSA we
must follow the evolution of the system until the cloud has grown by
several orders of magnitude.  In order to make the computational load
more manageable, we proceed as follows.  Starting at $t=0$, we
numerically evolve the system over the range $\bar r < r < r_{\rm
max}$, where $r_{\rm max} = 1000a_0$.  At $r=\bar r$, we require that
$g$ match onto the MSA solution.  To fix the behavior at the other end
of the interval, we use an expansion such as that given in
Eq.~(\ref{largeRexp}) to obtain an analytic approximation to $g$ for
$r_{\rm max} < r < \infty$.  We then require that the numerical
solution match onto this analytic approximation at $r=r_{\rm max}$.
As time goes on and the cloud expands, the spatial range of the
numerical integration must be enlarged.  To do this, we define a
position $r_{1/2}(t)$ by the condition $g(r_{1/2}, t) = 0.5$.  When
$r_{1/2}$ grows to twice its original size, we replace $r_{\rm max}$
by $r'_{\rm max} = 2r_{\rm max}$ and at the same time double the step
sizes in our integrations.  The new initial data is given for $\bar r
< r <r_{\rm max}$ by the previous numerical results and for $r_{\rm
max} < r < r'_{\rm max}$ by the large-distance analytic approximation.
We continue in this fashion, doubling the interval and step sizes as
required.

\begin{figure}[hbtp]\begin{center}
\PSbox{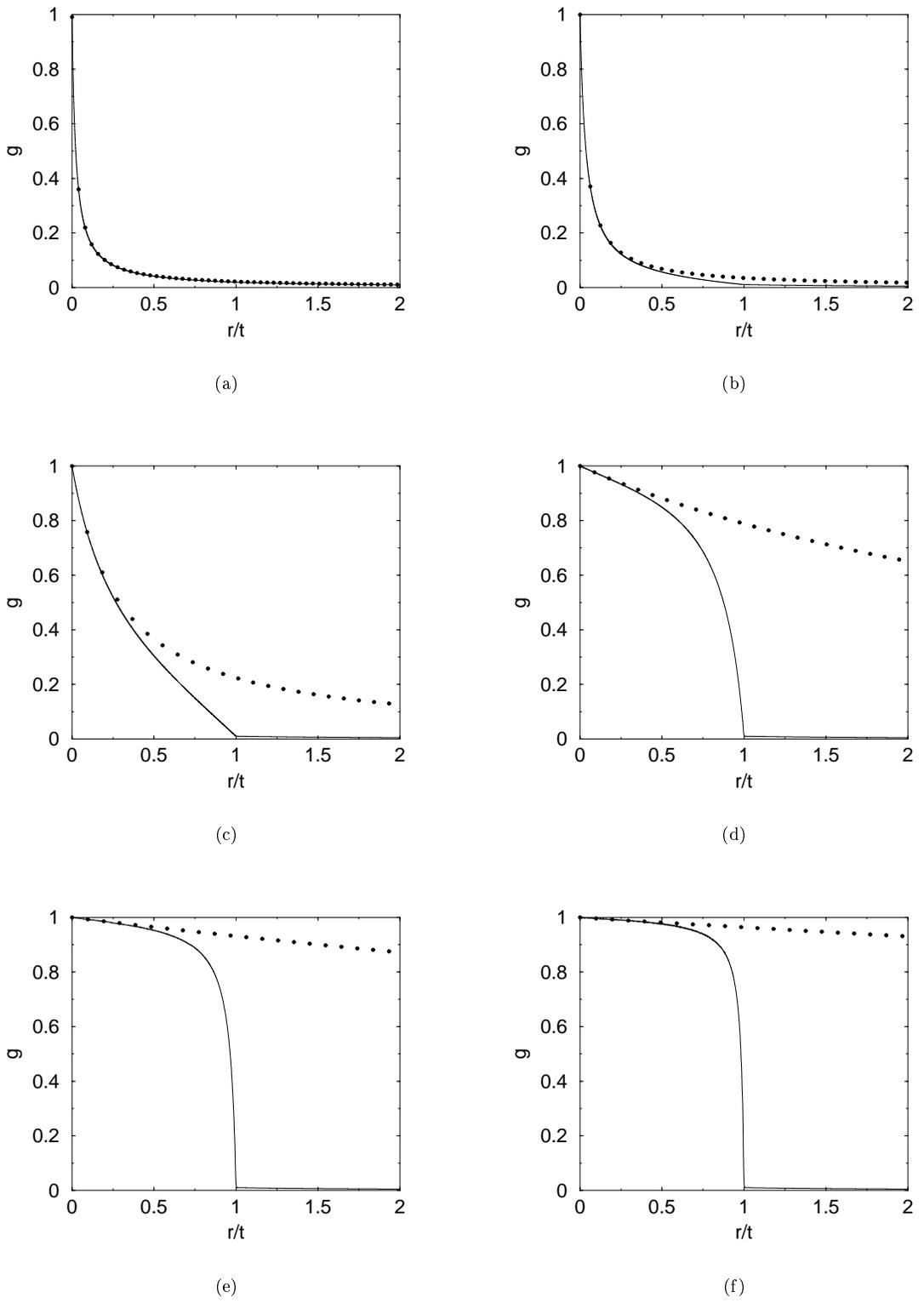
hscale=100 vscale=100  hoffset=-180    
voffset=-80}{3.5truein}{6.5truein}
\end{center} 
\medskip
\caption{Plots of the numerical solution (solid line) and MSA
approximation (dotted line) for $g(r,t)$ at various times.  Note that
the distance scale varies from panel to panel.  The panels correspond
to (a) $t=101 a_0$, (b) $t=1030 a_0$, (c) $t=1.11 \times 10^4 a_0$,
(d) $t=1.49 \times 10^5 a_0$, (e) $t=5.45 \times 10^5 a_0$, and (f)
$t=1.07 \times 10^6 a_0$. }
\label{fig:profiles}
\end{figure}

In Fig.~\ref{fig:profiles} we show our results for $g(r)$ at six
different times.  For comparison, in each case we also show the
predictions of the MSA.  (Note that the distance scale changes from
one panel to the next.)  For the two earliest times, $t=101a_0$ and
$t=1030a_0$, the MSA is clearly a good approximation to the exact
solution.  The only noticeable departure from the MSA appears to be
due to the retardation effects at large distance; indeed, note the
sharp bend in the data at $r=t$ that is visible at $t=1030a_0$ and all
later times.  The third time shown is approximately $t_{\rm crit}/2$.
At this point the departures from the MSA are beginning to be
significant, although the MSA still gives a fairly good qualitative
picture.  The times in the last three panels are all much greater than
$t_{\rm crit}$.  In these we see the numerical solution departing from
the MSA and evolving toward a step-like profile in which $g$ falls
from $1$ to 0 in a short interval near $r=t$.  This step is
essentially the radiation transformed into a sharp wavefront moving
outward at the speed of light; it arises because the cloud that is
the source of the radiation is now itself expanding at the speed of
light.  Although the step appears to be becoming narrower with
increasing $t$, this is due to the change in distance scale from one
panel to the next.  Indeed, we see from Eq.~(\ref{geq}) that far from
the origin $g$ is a function only of $r-t$, so that both the
shape and the width of the step are constant at large time.

\begin{figure}[bhtp]\begin{center}
\PSbox{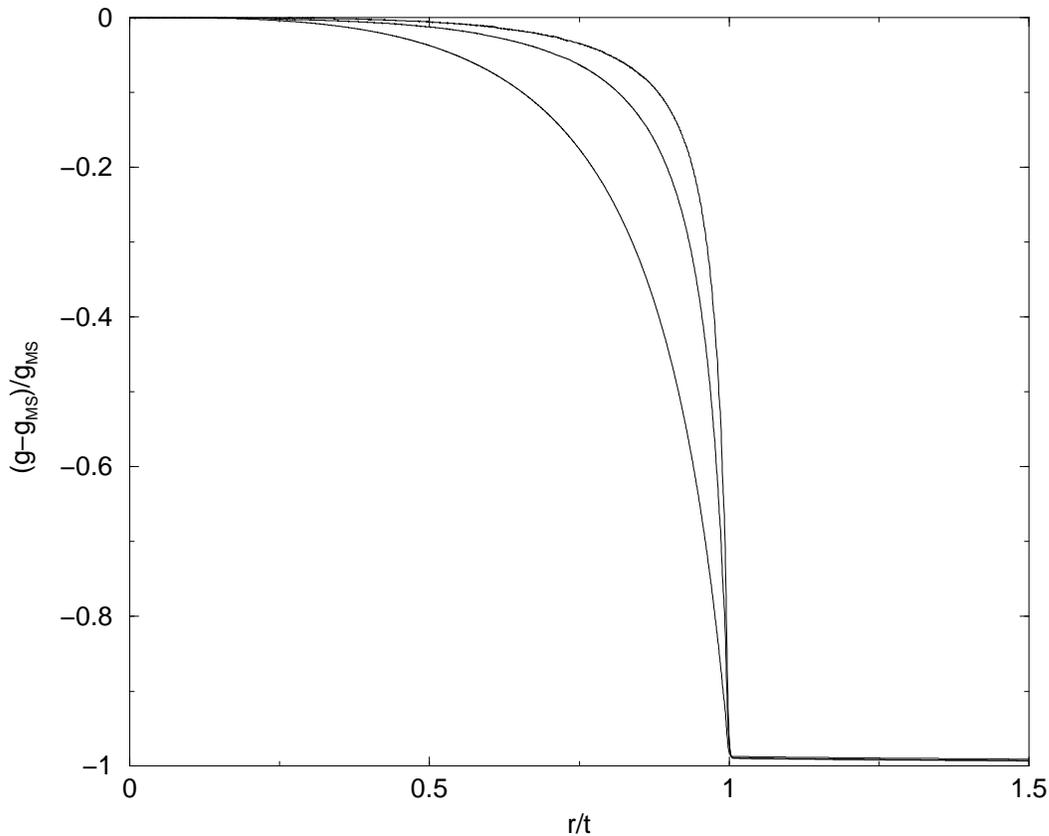 angle=-90
hscale=70 vscale=70  hoffset=-35    
voffset=400}{6.0truein}{4.5truein}
\end{center} 
\medskip
\caption{Fractional deviation of $g(r)$ from the MSA prediction.  From
inner to outer, the curves correspond to $t=1.49 \times 10^5 a_0$, $5.45
\times 10^5 a_0$, and $1.07 \times 10^6 a_0$. Note that the distance scale 
increases with time. }
\label{fig:deltaG}
\end{figure}

Despite this clear breakdown of the MSA at large distances, we find
that it remains a good approximation to the data at shorter distances.
To illustrate this, in Fig.~\ref{fig:deltaG} we plot the fractional
deviation of $g$ from the MSA prediction as a function of $r/t$ for
the times corresponding to the last three panels of
Fig.~\ref{fig:profiles}.  For any fixed $r/t < 1$ this fractional
deviation decreases with time.

\begin{figure}[hbtp]\begin{center}
\PSbox{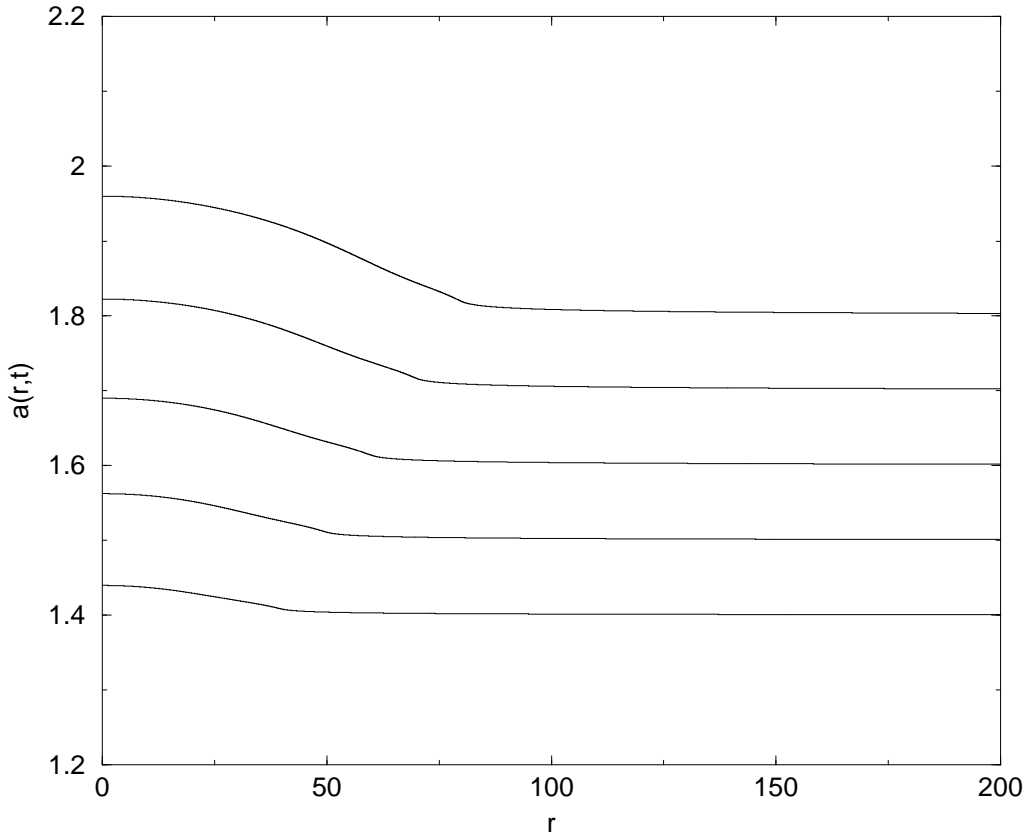 angle=-90
hscale=70 vscale=70  hoffset=-35    
voffset=400}{6.0truein}{4.5truein}
\end{center} 
\medskip
\caption{Plot of $a(r,t)$. Reading from the bottom, the curves
correspond to $t=40 a_0$, $50 a_0$, $60 a_0$, $70 a_0$, and $80 a_0$.
The distances are given in units of $a_0$.}
\label{fig:AofRsmallT}
\end{figure}

Another way to measure the deviation from the MSA is to define an
$r$-dependent cloud parameter $a(r,t)$ by substituting the actual
$g(r,t)$ into Eq.~(\ref{gm}).  In other words, 
\begin{equation}
    a(r,t) = {r g(r,t) \over 1 - g(r,t)} \, .
\end{equation}
If the MSA were exact, $a(r,t)$ would be independent of $r$ and equal
to $a_{MSA}(t)$.  In Fig.~\ref{fig:AofRsmallT} we plot $a(r,t)$ for
several times ranging from $40a_0$ to $80a_0$.  At all these times
$a(r,t)$ is relatively flat near the origin, gradually falls, and then
becomes approximately constant for $r>t$.  The increase in the actual
value in the latter region is very close to being linear in time, as
predicted by Eq.~(\ref{largeRg}).

\begin{figure}[thbp]\begin{center}
\PSbox{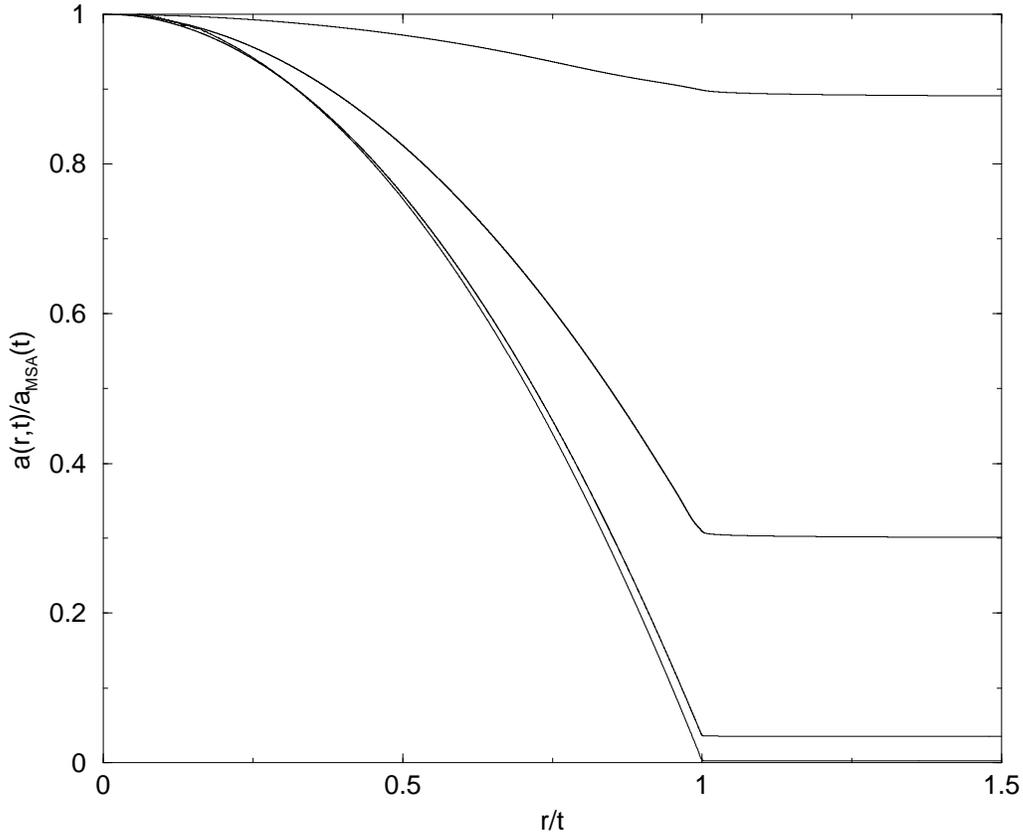 angle=-90
hscale=70 vscale=70  hoffset=-35    
voffset=370}{6.0truein}{3.6truein}
\end{center} 
\medskip
\caption{Plot of $a(r,t)$. Reading from the top, the curves
correspond to $t=101 a_0$, $1.03 \times 10^3 a_0$, $1.1 \times 10^4 a_0$, 
and $1.49 \times 10^5 a_0$.  Note that the distance scale increases with 
time.}   
\label{fig:AofRlargeT}
\end{figure}

We show the corresponding data for some larger times in
Fig.~\ref{fig:AofRlargeT},
although with a rescaling of axes to allow easier comparisons between
the data at the four different times.  Again we see that $a(r,t)$
is roughly constant at smaller $r$, varying by no more than 10\% for
$r<t/4$.  

\section{Concluding remarks}
 
In this paper we have investigated the reliability of the MSA when
applied to BPS monopoles in a theory with an unbroken non-Abelian
symmetry.  Whenever there are massless particles, the existence of
non-BPS states arbitrarily close in energy to the BPS solutions raises
the possibility that, no matter how small the kinetic energy, a
time-dependent solution will deviate from the moduli space of static
solutions to a sufficient extent to invalidate the MSA.  For the case of
Abelian massless fields, previous investigations have shown that the
radiation processes that might give rise to such departures fall as a power
of velocity and so can be neglected to leading approximation at low
energy.  We have found that this is not the case when the massless
fields are non-Abelian.  

The time-dependent fields in the cloud region surrounding the massive
monopoles act as sources for non-Abelian radiation.  As in the Abelian
case, radiation is suppressed at low velocities; in particular, the
power radiated is proportional to the square of the cloud kinetic
energy.  The crucial difference lies in the duration of the radiation.
The Abelian radiation in a multimonopole system falls off as the
massive monopoles recede from each other.  By contrast, the
non-Abelian radiation in our system continues unabated as the cloud
expands.  The net effect is that the field profiles depart sharply
from the MSA prediction by the time that the cloud velocity approaches
unity.  

Despite this, we do not find a complete breakdown of the MSA.  It
remains quite reliable as long as the cloud velocity is small.  Even
when the cloud has become relativistic, the MSA gives a good
description of the fields well inside the cloud (say for $r \lesssim
t/4$).  If the massive monopoles within the cloud remain
nonrelativistic, they will always be in the region 
where the MSA is valid.  Hence, one should be able to use the
MSA to describe both the motion of the massive monopoles and the
transfer of energy between the massive and massless monopoles.
Furthermore, the cloud does survive as a clearly identifiable
object in the classical solution, although the smoothly varying
profile of the static BPS solution is transformed into a relativistic
wavefront.  In a sense, the modifications to the MSA that we have
found are the minimal ones consistent with the relativistic
bound on velocity.

Although we have focussed on the case where there is a strictly
massless monopole, we should make a few remarks on the situation when
one of monopoles is massive but much lighter than all the others.  For
the analogue of our SO(5) example Eq.~(\ref{MSAenergy}) is replaced by 
\begin{equation}
   E = {2\pi \over e^2 r_0} \left({r + r_0 \over r}\right) \dot r^2
      = {1\over 2} \mu \left({r + r_0 \over r}\right) \dot r^2 \, ,
\end{equation} 
where $r$ is the separation of the two monopoles, $\mu$ is the
reduced mass of the monopole pair, and $r_0 \equiv 4\pi/e^2\mu$
gives the size of the lighter monopole core when the monopoles
are well separated.

When $r \lesssim r_0$, the two monopoles are overlapping, with the core
of the lighter monopole enclosing the heavier monopole.  This core is
then much smaller than its natural size and has a radius
approximately equal to $r$.  In this regime $\dot r$ grows linearly
with time, as in the massless case, and only tends to a constant value
$\sqrt{2E/\mu}$ after the two cores separate.  If $E \ll \mu$, the two
monopoles are always nonrelativistic, and the MSA is reliable.  If
instead $E \gtrsim \mu$, a breakdown of the MSA, similar to that which
we have seen in this paper, occurs while the two monopoles are still
overlapping.  We would expect this to lead to a deformation of the
core profile of the lighter monopole so that initially, like the cloud
in the massless case, it resembles a spherical wavefront expanding at
the speed of light.  However, when the core radius become of order
$r_0$, this wavefront presumably splits into a shell of radiation that
continues to expand and a massive core that is left behind.  The
heavier monopole will eventually separate from this core, but not from
the outgoing shell of radiation.

For this two-body system the validity of the MSA is assured if the
initial relative velocity is small, since the final configuration 
is nonrelativistic if the initial one is.  This need not be true in a
three-body system.  An example occurs in an SU(3) theory broken to
U(1)$\times$U(1) with one of the fundamental monopoles being much
lighter than the other.  Consider a configuration containing one of
the light and two of the heavy monopoles.  Because there can be a net
transfer of energy from the heavy to the light monopoles, it is
possible for the light monopole to initially have $v \ll 1$ but to
emerge with a relativistic final velocity.  As a result, the MSA fails
at large times, as was first pointed out by Irwin \cite{Irwin:2000rc}.

Finally, let us consider what happens as we move toward the
strong coupling (large $e$) limit, where the duality conjecture leads
us to expect the solitons and elementary particles to exchange roles.
Even without taking into account the quantum corrections to the
classical field profiles, we can see how the classical soliton picture
is lost.  For the massive monopoles this is signalled by the fact that
the core radius $\sim 1/(e^2 M)$ becomes less than the Compton
wavelength, so that a nonrelativistic monopole cannot be sufficiently
localized for its solitonic properties to be directly observable.  For
the massless monopoles, the transition can be seen by focussing on the
two phases in the evolution of the massless cloud.  For a time
interval of order $t_{\rm crit}\sim 1/(e^2 E)$ the cloud behaves like
a slowly expanding soliton, as predicted by the MSA.  Beyond this
time, it resembles a wave front expanding at the speed of light.  In
the strong coupling limit, $t_{\rm crit} E$ is much less than unity.
The uncertainty principle then implies that the period when the
massless monopoles behave like nonrelativistic solitons is
unobservably short; instead, they will always appear as a wavefront of
radiation moving at the speed of light.  In retrospect, this makes it
seem quite natural that, even in the weak-coupling regime, the cloud
cannot be kept nonrelativistic.

\acknowledgments 
This work was supported in part by the U.S. Department of Energy.

\end{document}